# Spin dynamics in semiconductor nanocrystals

J. A. Gupta and D. D. Awschalom[1]

*Department of Physics, University of California, Santa Barbara, CA 93106*

Al. L. Efros

*Naval Research Laboratory, Washington DC 20375*

A.V. Rodina[2]

*Solid State Physics Institute, Technical University of Berlin, D-10623 Berlin, Germany*

## Abstract

Time-resolved Faraday rotation is used to study both transverse and longitudinal spin relaxation in chemically-synthesized CdSe nanocrystals (NCs) 22-80Å in diameter. The precession of optically-injected spins in a transverse magnetic field occurs at distinct frequencies whose assignment to electron and exciton spins is developed through systematic studies of the size-dependence and theoretical calculations. It is shown that the transverse spin lifetime is limited by inhomogeneous dephasing to a degree that cannot be accounted for by the NC size distribution alone. Longitudinal spin relaxation in these NCs occurs on several distinct timescales ranging from 100ps-10μs and exhibits markedly different dependencies on temperature and field in comparison to transverse spin relaxation.

[1] To whom correspondence should be addressed: awsch@physics.ucsb.edu
[2] On leave from A.F. Ioffe Physico-Technical Institute

## I. INTRODUCTION

The size-tunable energy level spectrum of semiconductor quantum dots (QDs) has driven interest in applications such as QD-based lasers[1] and fluorescent labels for biological molecules.[2] Continuing interest in the dynamics of carrier spins localized within QDs is motivated by the emerging fields of semiconductor spintronics[3] and quantum computation.[4] Proposals for the latter hope to exploit expected increases in spin relaxation lifetimes due to fully three-dimensional confinement of electron spins. Certain channels of homogeneous spin-spin interactions may be absent in strongly-confining QDs as the discrete energy spectrum and the exclusion principle suppress the excitation of multiple electron-hole pairs within single QDs by circularly polarized, resonant photons. Recent studies have suggested that hyperfine interactions between electrons and nuclei may be the operative mechanism for carrier spin relaxation.[5]

Here we present a series of measurements that characterize both transverse and longitudinal spin dynamics in chemically-synthesized nanocrystal quantum dots. It was shown in an earlier paper that spin precession following optical excitation with circularly polarized light in a transverse magnetic field occurs at up to 4 distinct frequencies,[6] although assignment of the precession signatures to electrons, holes, or excitons was uncertain. Extension of the Faraday rotation technique to include independently-tunable pump and probe pulses has aided the clear observation of spin precession in the full range of samples studied. Calculations of the electron $g$-factor taking into account a size-dependent energy shift of allowable levels closely agree with the experimental data for one of the observed frequency components. Assignment of the other frequency component(s) to excitons remains uncertain; while the values obtained from the Faraday rotation are consistent with the exciton $g$-factor obtained from magnetoabsorption data, unambiguous interpretation is hindered by considerations of anisotropy present in these nanocrystals (NCs). Efforts to calculate the exciton $g$-factor for a randomly oriented ensemble of wurtzite NCs have led us to posit that exciton precession might only be exhibited in a special subset of ‘quasi-spherical’ NCs where anisotropy contributions from the crystal structure and physical shape cancel out.

Inhomogeneous dephasing limits the transverse lifetimes of both electron and exciton spins in these samples. By using measurements of the size-dependent electron $g$-factor to estimate the dephasing rate, we find that dephasing from the size distribution can not account for the observed rate in all but the smallest NCs. This suggests that contributions to electron dephasing from surface conditions[7] and/or NC anisotropy may also be important. Longitudinal spin relaxation in NCs was studied by applying a magnetic field along the observation direction (the Faraday geometry). The decay of spin polarization in this case is found to occur on ~100ps, 10ns and 10μs time scales. The microsecond decay time sensitively depends on the applied field and temperature in comparison to the shorter decay components, which exhibit more gradual trends. In addition to the monotonic decay of longitudinal polarization, spin precession is found to occur in the Faraday geometry, raising questions concerning the nature of optically excited states in these NCs.

The remainder of this paper is organized into four sections. In Sec. II we describe the set of samples that have been studied and the energy spectrum of allowed states. We also review the experimental methods of time-resolved Faraday rotation and absorption that are used to monitor carrier spin and population dynamics. Section III focuses on our efforts to assign the multiple precession signatures to electrons and excitons, while Sec. IV details our measurements of inhomogeneous dephasing. Studies of longitudinal spin relaxation are presented in Sec. V, and we conclude in Sec. VI. Further details of the quasi-spherical model for exciton spin precession are provided in an Appendix.

## II. SAMPLE INTRODUCTION AND EXPERIMENTAL METHODS

Chemically synthesized CdSe nanocrystals result from the pyrolytic reaction of organometallic precursors in a coordinating organic solvent.[8] Single-crystals with the wurtzite structure were studied ranging from 22-80Å in diameter, with a size distribution of ~5-15%. The NCs are slightly prolate (elongated along the c-axis); with a size-dependent mean aspect ratio (1+*m*) that may range from 1.0 to 1.3 with increasing size.[8b] The 80Å NCs are somewhat more prolate, with an aspect ratio of ~2. Although the distribution of aspect ratios is not well characterized, a variation of ± 0.2 may be expected for samples 22-57Å in diameter.[9] Surface-modified CdSe/CdS core/shell NCs were also studied; the epitaxial growth of a higher bandgap shell is known to reduce the efficiency of carrier trapping by surface states, providing improved luminescence quantum yields up to 90%.[10] Samples for the optical measurements were prepared by dissolving the NCs in an organic polymer which freezes into an optically transparent glass at cryogenic temperatures.

In analogy to atomic Hydrogen, the discrete spectrum of states in NCs results from an approximately spherical confinement potential provided by the bandgap contrast between the semiconductor and insulating matrix. The net wavefunction of excitons occupying these states is described by the product of electron and hole wavefunctions comprised of parts related to the intrinsic unit cell of the solid and an envelope function that satisfies boundary conditions imposed by the NCs' finite size. As an example, the ground exciton state is denoted $1S_{3/2}1S_e$ where the term $1S$ refers to the hydrogenic shell state and the subscripts refer to either $J$=3/2 holes or $S$=1/2 electrons.[11,12] Ground and excited states are visible in the linear absorption spectrum of samples with relatively good size distributions as inhomogeneously broadened peaks [Fig. 1(a)]. In perfectly spherical NCs with cubic crystal structure, the $1S_{3/2}1S_e$ ground exciton state is eightfold degenerate, comprising the possible combinations of hole and electron spins. This degeneracy can be broken in CdSe NCs resulting in a multiplet of five 'fine structure' states whose relative order and spacing depends on the wurtzite crystal structure, prolate shape and e-h exchange interaction [Fig. 1(b)]. Although these states can not be directly resolved in absorption spectra, they have been observed using size-selective techniques to circumvent broadening due to the size distribution.[13] The states are labeled by their angular momentum projection onto the crystal c-axis (0,±1 or ±2); the superscripts '$U$' or '$L$' refer to higher and lower energy states with the same projection.[13,14] In NCs whose c-axis is oriented along the observation direction, optical selection rules are well defined and dictate that circularly polarized $\sigma^{\pm}$ light will couple to the ±1 states respectively. Selection rules are poorly defined in the large majority of NCs whose c-axis is oriented at a random angle; superpositions of both states may be excited in such NCs. Previous work suggests that radiative recombination of optically-excited carriers occurs upon relaxation to the ±2 'dark exciton' states;[15] the timescale for such relaxation is not well known but may be of order ~200ps.[16]

Pump-probe techniques were employed to monitor carrier population and spin dynamics in the time domain. Synchronized optical parametric amplifiers with a repetition rate of 250kHz produce ~150fs pump and probe pulses with energies $E_{pump}$ and $E_{probe}$ that are independently tunable across the visible spectrum. A pulsed white light continuum is also produced that can alternatively serve as a probe pulse. The pump and probe pulses are normally incident on the sample and focused to a ~100μm spot using an achromatic doublet lens. Conventional mechanical delay lines control the relative time delay, t, between pump and probe pulses and optical choppers are used to modulate both pulse trains, enabling the use of lock-in amplifiers. The sample sits in the bore of a magneto-optical cryostat capable of magnetic fields up to H=7T and temperatures from T=2-300K.

Carrier population dynamics were monitored through time-resolved absorption (TRA) as a pump-induced change in the absorption spectrum detected at an energy E:

$$\Delta a\,(E,t)L \equiv \left(a_0 - a\,(t)\right)L = \ln\left(\frac{T(t)-T_0}{T_0}+1\right). \qquad (1)$$

Here $L$ is the sample thickness and $\alpha$ ($\alpha_0$) is the semiconductor absorption spectrum in the presence (absence) of the pump. In practice, $\Delta\alpha L$ is computed from the corresponding transmitted probe intensities, $T(t)$ and $T_0$; positive values typically represent state-filling effects where the probe is preferentially transmitted due to the occupation of states by pump-excited carriers that would otherwise be available for absorption of probe photons. The white light continuum serves as the probe in these experiments, facilitating measurement of $\Delta\alpha L$ across the entire NC absorption spectrum. Changes in the transmitted probe intensity are detected using either a photomultiplier tube or photodiode array which are attached to a 0.5m spectrometer.[17] Both pump and probe pulses are circularly polarized, with a relative helicity that can be controlled using variable retarders.

Carrier spin dynamics are monitored with the technique of time-resolved Faraday rotation (TRFR).[18] The circularly polarized pump pulse typically excites a population of electron-hole pairs that are spin polarized along the optical path. The resultant spin magnetization and spin-selective state-filling effects produce a circular birefringence in the sample whose component along the optical path is detected as the rotation of a linearly-polarized probe pulse. Sensitivity to rotation angles > 10μrad is achieved using a balanced photodiode bridge. If the magnetic field is applied along the observation direction (Faraday geometry), the pump-excited spins comprise a highly non-equilibrium population whose relaxation involves energy transfer to the lattice via phonons. In the simplest case, such relaxation follows an exponential decay with a time constant $T_1$. When the magnetic field is applied perpendicular to the observation direction (Voigt geometry), the pump-excited spins comprise a coherent superposition of states whose temporal evolution corresponds to Larmor spin precession at a frequency $n_L = g m_B H / h$ proportional to the spin $g$-factor ($m_B$ is the Bohr magneton). Transverse spin relaxation yields a lifetime $T_2^*$ which may contain contributions from longitudinal and purely transverse spin relaxation, as well as inhomogeneous dephasing.

## III. MULTIPLE PRECESSION FREQUENCIES

Figures 2(a)-(d) show characteristic spin precession in the Voigt geometry for CdSe NCs 25-80Å in diameter. It has previously been shown that low-field spin lifetimes in these samples are ~3ns;[6] the ~50ps lifetimes here are a result of inhomogeneous dephasing, discussed in Sect. IV. Data are usually taken with the pump and probe energies in the vicinity of the $1S_{3/2}1S_e$ exciton state. The visibility of the spin precession oscillations can depend sensitively on the choice of detection energy. This is illustrated in Fig. 2(e) for two choices of $E_{probe}$ near the ground state in 57Å NCs (2.05eV). Because the Faraday rotation signal is proportional to the difference in refractive indices for the right- and left-circularly polarized components of the linearly polarized probe,[19] the magnitude and sign can change with $E_{probe}$ in the vicinity of an absorption resonance. The ability to independently tune pump and probe energies has proven useful in maximizing the prominence of spin precession, particularly in the smaller NCs < 40Å in diameter, where initial experiments reported only marginal results.[6] It is often the case that there is a monotonically-decaying background contribution to the Faraday rotation signal in this geometry. A possible origin of this signal will be discussed in the context of the Faraday geometry measurements in Section V.

Modulation of the oscillation envelope indicates the presence of multiple frequency components that are directly resolved in a fast Fourier transform (FFT) of the data [Fig. 2(f)]. FFTs of spin precession over a broad range in detection energy reveal no variation in spectral content; only the amplitude of the oscillatory signal is affected. Precise measurement of the *g*-factors associated with these components can be obtained by plotting the positions of peaks in the FFT spectrum versus magnetic field. Figure 3 and Table I show a compilation of *g*-factors measured in our sample set. Two *g*-factors have been observed in NCs ranging from 25-80Å, with the notable exception of the 57Å NC sample, where four *g*-factors were visible.

We assign the lower frequency precession to the electron *g*-factor which is considerably shifted from the bulk value of $g_e$=0.68 in CdSe.[20] This shift and the trend with size can be qualitatively understood by modifying equations for $g_e$ in bulk semiconductors[21] to account for the confinement-induced shift of electron energy within the conduction band:[22]

$$g_e = g_0 - \frac{2}{3}\frac{E_P \Delta_{SO}}{(E_g + E + \Delta_{SO})(E_g + E)} . \qquad (2)$$

Here $g_o$~2 is the free electron $g$-factor, $E$ is the electron energy measured from the bottom of the bulk conduction band, $E_g$ is the semiconductor bandgap, $E_p$ is the Kane energy parameter and $D_{so}$ is the spin-orbit splitting of the valence band. The electron $g$-factor implicitly depends on the NC radius $R$ through the size-dependent electron energy which can be written as $E\sim a/R^2$ in the simplest parabolic approximation.[14] The solid curve in Fig. 3 shows a calculation of $g_e$ from Eq. (2) which treats $a$=267 eVÅ$^2$ as the only fitting parameter and uses literature values of $E_g$=1.839eV (for bulk CdSe at 10K)[12], $D_{so}$=0.42eV [12] and $E_p$=19.1eV.[23] The dotted line was calculated using a better description of the size-dependent electron energy and consideration of finite boundary conditions at the NC surface.[7(b)] It has been shown that the latter may result in changes to calculated vales of $g_e$ by ~5-10% in NCs;[7] a variable characterizing this contribution was treated as the only fitting parameter in the calculations. Our assignment of this precession component to electron spins is based on the good agreement between the experimental data and these calculations. Higher values of $g_e$ found in the CdSe/CdS core/shell NCs (Table I) are most likely due to electron penetration into the CdS shell, as (bulk) CdS has a larger $g_e$ value of 1.76.[24]

As a first step toward assignment of the second precession component to exciton spins, we have performed magnetoabsorption experiments in the Faraday geometry that should be simply connected with the band-edge exciton spin splitting.[25] Gaussian peaks are fit to inhomogeneously broadened absorption spectra [c.f. Fig. 1(a)] taken with $s^+$ and $s^-$ light helicities. The splitting between spin-up and spin-down exciton states was extracted from the center positions of the ground state exciton peak and plotted versus field.[26] Large fields (H>3T) were needed in order to observe the spin splitting (<1meV), which is much smaller than the inhomogeneous absorption linewidth (~50meV). Included in Fig. 3 are values of $g_{exc}$ calculated from magnetoabsorption data for 40Å and 57Å NCs. Within experimental error, the values of $g_{exc}$ obtained from this procedure are consistent with the larger $g$-factor components measured from Faraday rotation data.

Assignment of the $g$-factors $g_{2-4}$ to exciton precession is complicated by two factors:

*(i)* Random orientation of NCs in the plastic films

*(ii)* The multiplet of possible states within the ground state exciton.[14]

The random orientation is a problem because the exciton $g$-factor is expected to be strongly anisotropic in CdSe NCs due to the non-spherical shape and wurtzite crystal structure.[24] Thus, measured values of the exciton $g$-factor for individual NCs can depend on the orientation of the $c$-axis relative to the applied field and observation directions. For our ensemble measurements, this anisotropy must be averaged over a random distribution. Furthermore, the shape, crystal structure and exchange interaction in NCs result in a multiplet of fine structure states whose relative spacing *and ordering* can depend on the exact size and shape of the NC [Figs. 4(a)-(b)]. We expect that the variation in aspect ratio present within our samples is large enough that different NCs within the ensemble can have a significantly different arrangement of these fine structure levels. Depending on how quickly (or if) relaxation occurs from optically excited states to the lowest energy state, different exciton $g$-factors may be obtained because each of these levels will exhibit a characteristic spin splitting with magnetic field. Association of the larger $g$-factor component with exciton states provides the possibility for explaining the multiple $g$-factors $g_2$-$g_4$ observed in the 57Å NCs (c.f. Table I) in terms of the fine structure levels, although it is unclear why this is the only size in which $g_3$ and $g_4$ could be resolved.

It should be noted that electron and exciton precession mustn't necessarily occur together within the same NC. It is possible that exciton precession is only exhibited in a special subset of NCs, while electron precession is expressed by the majority of NCs in the ensemble. Because our efforts to average an anisotropic exciton $g$-factor didn't qualitatively agree with the experimental data,[27] we speculate that exciton spin precession may only occur within 'quasi-spherical' NCs where the anisotropy due to the physical shape exactly cancels that due to the wurtzite crystal structure. Such cancellation is possible because the two terms have opposite sign in CdSe NCs.[14] For a given NC diameter, a choice of aspect ratio within the range expected for the ensemble can be found where the anisotropy due to shape and crystal structure offset. The exciton $g$-factor in this subset of NCs may be perfectly isotropic thus simplifying the interpretation by obviating the averaging of energy level structure and anisotropy over the ensemble that would otherwise be necessary. The dotted line in Fig. 3 shows a calculation of the ground state exciton $g$-factor from the spin splitting of $\pm 1^L$ fine structure levels in quasi-spherical NCs using the relation:

$$g_{exc} = (g_e - 3g_h)/2\,. \qquad (3)$$

The hole $g$-factor was treated as a size-independent fitting parameter; all size-dependence was assumed to come from $g_e$. Details of this calculation will be discussed in Appendix I; while the agreement is promising, a definitive assignment of the higher $g$-factor component to exciton precession requires further study.

Exciton spin precession with a ~50ps spin lifetime has previously only been observed in thin GaAs/AlGaAs quantum wells under stringent conditions of low-energy excitation and low temperature which stem from the sensitivity of the hole spin to energy relaxation and thermal processes.[28] If our assignment of higher-frequency precession to excitons is correct, the dependencies of exciton spin lifetimes on temperature and excitation energy are completely different in NCs. It was shown earlier that both electron and exciton precession persist to room temperature with ns-scale lifetimes at low field.[6] Here we will also show that spin relaxation in these NCs is surprisingly indifferent to energy relaxation, even when initial excitation occurs hundreds of meV above the ground state exciton energy.

To monitor possible changes in carrier population dynamics with excitation energy, Figure 5 shows TRA data taken at three different pump energies, indicated by the arrows in Fig. 5(a). Positive peaks in Figs. 5(b)-(d) occur due to state-filling effects and correspond to the position of the $1S_{3/2}1S_e$ and $2S_{3/2}1S_e$ states estimated from the absorption spectrum [c.f. Fig. 1(a)]. Even though the carrier population resides in the ground state, absorption of probe photons at the $2S_{3/2}1S_e$ states' energy is suppressed due to the shared $1S_e$ level.[29] Because the choice of excitation energy in Fig. 5(b) selectively excites larger NCs in the size distribution, the state-filling peaks are redshifted relative to Figs. 5(c)-(d). The difference in $\Delta\alpha L$ with probe polarization evidenced by data taken at t=20ps indicates that the population of spin-polarized carriers is a small fraction of the total carrier population. In principle, the maximum spin polarization of 100% can be achieved in bulk CdSe for near-resonant excitation along the $c$-axis; the small amount here probably reflects the randomly oriented ensemble of NCs in these samples. It is interesting to note that the TRA signal persists over the entire repetition interval of 4μs, indicating that carriers populate NCs on this timescale. This may be consistent with the ~1 μs decay times measured from time-resolved photoluminescence experiments,[15,30] although evidence is shown in Sect. V that spin-polarization can last for even longer times. The negative feature which tracks the excitation energy is not well understood; the lack of polarization and longevity of the signal might be signatures of spectral-hole burning, although such signals are generally positive.[31] As established in previous studies,[32] energy relaxation down to the $1S_{3/2}1S_e$ state occurs quickly (<1ps), accounting for the similar $\Delta\alpha L$ signals in Figs. 5(c)-(d) and directly verified in Fig. 5(e) by the fast rise time (~500fs) of the state-filling signal with high $E_{pump}$. Figure 5(f) shows that the initial decay of $DaL$ due to carrier recombination is independent of temperature up to the measured 70K. The longer decay component is ~12ns, indicating that the ~3ns transverse spin lifetimes in these NCs at low field are not limited by carrier recombination.

Figure 6 shows complementary Faraday rotation data taken over the same range of excitation energy. Spin precession lifetimes are roughly constant over this range of $E_{pump}$, indicating that spin coherence is preserved during the rapid energy relaxation process, even when initial excitation occurs into higher exciton states within the NC (i.e. top panel). This surprising result holds despite the complex spectrum of states accessible to higher energy pump photons. In addition to the fine structure levels present within the ground $1S_{3/2}1S_e$ exciton state, each of the higher excited states has its own characteristic multiplet of fine structure levels. The overlap of these levels for different NCs within the ensemble makes the detailed treatment of energy and spin relaxation in these NCs intractable. FFTs of the spin precession data indicate that while the exciton spin lifetime is largely unchanged, the relative spectral weight of exciton precession decreases with decreasing $E_{pump}$, a trend that has also been observed in the 80Å and 40Å NCs. Speculations on this effect in the context of the 'quasi-spherical' model will be discussed in the Appendix. For now, we note that this trend reinforces the notion that exciton and electron spin precession may be exhibited by different NCs within the ensemble.

## IV. INHOMOGENEOUS DEPHASING

Inhomogeneous dephasing occurs when an ensemble of spins precess at a continuous distribution of frequencies. Dephasing leads to an artificially short decay time, even though the individual spins may still have a well-defined phase relative to their initial direction. In conventional nuclear magnetic resonance, dephasing commonly results from inhomogeniety in the external magnetic field. Even small variations in field suffice to limit measured transverse spin lifetimes because intrinsic nuclear spin lifetimes are long and the samples are large. For the electron spin lifetimes and small sample volume measured in these optical experiments, inhomogeniety of the applied field can be neglected. The dephasing of spin precession in these NCs is caused by intrinsic sample inhomogeniety, the most straightforward example of which is the size distribution.

As demonstrated in Fig. 3, both electron and exciton $g$-factors are dependent on the NC diameter, so that the sample size distribution will result in a corresponding distribution in $g$-factors. To estimate how the resultant inhomogeneous dephasing affects the ensemble spin magnetization, we start by assuming a Gaussian distribution of $g$-factors about over $N$ spins:

$$N(g) = \frac{N}{\sqrt{2p}\,\Delta_g} \exp\left[-(g - g_0)^2 / 2\Delta_g^2\right]. \tag{4}$$

Here $D_g$ is the standard deviation of the distribution centered about $g_o$. The net magnetization at each value of $g$ is:

$$M(t,g) = N(g)\exp(-t/T_2^*)\cos(g m_B H t/\hbar), \tag{5}$$

so that the ensemble magnetization is obtained by integrating $M(t,g)$ over all possible $g$:

$$M(t,g_0) = \frac{N}{\sqrt{2p\Delta_g}} \int_{-\infty}^{\infty} e^{-(g-g_0)^2/2\Delta_g^2} e^{-t/T_2^*} \cos(g m_B H t/\hbar) dg$$
$$\downarrow \tag{6}$$
$$M(t,g_0) = N e^{-t/T_2^*} \exp\left(-t^2/2(\hbar/m_B H\Delta_g)^2\right)\cos(g_0 m_B H t/\hbar).$$

When dephasing dominates the spin dynamics, the precession envelope acquires a Gaussian shape with a standard deviation $D_t$ that is inversely proportional to the applied field, defining an 'inhomogeneous' lifetime:

$$T_{inh} \equiv \Delta_t = \frac{\hbar}{m_B H \Delta_g}. \tag{7}$$

In practice it is easier to measure the frequency linewidth of the Fourier transform (FT) since it is tedious to fit the complicated spin precession dynamics in the time domain. While the FT of spin precession data in the absence of dephasing follows a Lorentzian function if $T_2^*$ is well defined, the FT under dephasing due to a Gaussian distribution of $g$-factors is Gaussian with a standard deviation in frequency:

$$\Delta_f = (2p\Delta_t)^{-1} = \frac{m_B H \Delta_g}{h}. \tag{8}$$

The deviation $D_g$ can thus be calculated from a plot of $D_f$ versus H.[33]

In order to quantify dephasing rates in these NCs, we apply this model of dephasing for both electron and exciton precession using FFTs of spin precession data, shown in the inset of Fig. 7(a). The two peaks in the FFT consistently resolvable in the sample set were fit with Gaussian functions to extract the linewidths. Note that although the central regions of each peak are reasonably well fit, this procedure is only an approximation because the complete lineshapes are neither purely Gaussian nor Lorentzian. Results are shown in Figs. 7(a)-(b) for electron and exciton components in 80Å and 40Å NCs. A priori, the linear relationship in these plots suggests that dephasing in these samples does result from a Gaussian distribution, centered about the nominal electron and exciton $g$-factors. In nearly all of the samples studied,[34] the dephasing rate associated with the exciton $g$-factor component is larger. This cannot be explained if exciton dephasing arises solely from the size distribution since $g_{exc}$ depends less sensitively on size than does $g_e$ [Fig. 3]. If the multiple exciton-like $g$-factors resolved in 57Å NCs are present but unresolved in the other NC samples then there may be a contribution to the exciton linewidth from the field-dependent splitting of the triplet that could account for at least part of the discrepancy.

When compared to core-only NCs, the electron dephasing rate is significantly higher in core/shell NCs [Fig. 7(b)] which may be attributed to fluctuations in shell thickness and thus $g_e$ among the ensemble. Figure 7(c) shows that electron dephasing in core-only NCs decreases with size. This also is unexpected from the size-dependence of $g_e$ if one assumes a constant size distribution in each of the samples because the slope of the curve in Fig. 2 increases with decreasing size. Although the exact size distribution in these samples is not well characterized, it is expected to fall between ~5-10% for 25-57Å NCs and ~15% for 80Å NCs.[35] We estimate the standard deviation in $g_e$ from the size distribution by assuming a Gaussian deviation of 5% which can be converted into a deviation $D_{g\text{-}size}$ using the data in Fig. 3 as a calibration curve. This is compared to the values $D_{g\text{-}exp}$ directly obtained from spin precession data using Eq. (8). The inset to Fig. 7(c) shows that $D_{g\text{-}size} << D_{g\text{-}exp}$ for all but the smallest NCs studied, suggesting that electron spin dephasing in these samples is not solely due to the size distribution. Incidentally, a 10% size distribution results in a dephasing rate that is too high for the smaller NCs (left triangles in the inset), however the choice of size distribution should be subject to the constraint that $D_{g\text{-}size} = D_{g\text{-}exp}$. No reasonable choice of size distribution can account for the observed dephasing rate in the 80Å QDs; $D_{g\text{-}size}$ for the expected 15% size distribution in this sample is shown as the right triangle in the inset. An additional contribution to the observed dephasing rate may come from the wurtzite crystal structure; previous calculations for bulk semiconductors suggest that the crystal-field splitting of heavy-and light-hole valence bands by an energy $D_{cr}$ results in a slight anisotropy of $g_e$ relative to the $c$-axis.[36] Using expressions for $g_e^{\|}$ and $g_e^{\perp}$ in Ref. 36, it can be shown that if one substitutes $E_g+E$ for $E_g$ as was done in Eq. (2) and uses the approximation $\Delta_{so} >> \Delta_{cr}$ that is valid for the bulk CdSe ($\Delta_{so} \sim 420meV, \Delta_{cr} \sim 25meV$) this anisotropy:

$$\Delta g_e = g_e^{\perp} - g_e^{\|} \approx \frac{E_p \Delta_{cr}}{2(E_g + E)(E_g + E + \Delta_{cr})} \quad (9)$$

decreases with increasing energy in the conduction band. Thus the dependence of electron energy on quantum confinement will lead to a decrease in anisotropy and its contribution to dephasing with decreasing size, in qualitative agreement with the observed trends. However, one has to also account for valence band splitting due to the non-spherical NC shape, so that $Dg_e$ must be averaged over an ensemble of sizes and shapes before its contribution to electron dephasing can be estimated. As a rough indication of scale, we estimate that (without averaging) $Dg_e$ can be of order 1-5% for a typical NC. Finally, the calculations in Ref. 7 indicate that a comparable change in $g_e$ can be obtained with modest adjustment of parameters that characterize the region at the surface over which the electron wave function goes toward zero. Further studies beyond the scope of this paper are required to average the anisotropy and surface effects in order to calculate the possible contributions to dephasing in these NCs.

## V. LONGITUDINAL SPIN RELAXATION

As discussed in Section II, measurements of spin dynamics in the Faraday geometry can be used to quantify longitudinal spin relaxation which in the simplest cases proceeds with a decay time $T_1$. Such relaxation results from the exchange of energy between a spin population and the host lattice as the system seeks to achieve a thermal equilibrium value of spin polarization. Figures 8(a)-(b) show the dependence of Faraday rotation data taken in the Faraday geometry as a function of magnetic field and temperature. Data here were fit with a bi-exponential decay plus an offset: $A\exp(-t/T_1^a) + B\exp(-t/T_1^b) + C$, yielding spin lifetimes plotted in Figs. 8(c)-(d). As before, this should be viewed as a convenient parametrization of the data since the true decay of spin magnetization in these samples is more likely to be non-exponential. While the ~100ps-scale decay component decreases with field, the 10ns-scale component dramatically increases over a surprisingly small range in magnetic field of < 0.01T before settling down to ~8ns at H=1T. This dependence on field is in distinct contrast to the proportionate reduction of transverse spin lifetimes by inhomogeneous dephasing. Although an increase in $T_1$ is consistent with simple expectations based on Boltzmann statistics, the field range over which this occurs corresponds to an energy scale (~6μeV) that is much smaller than $k_BT$ (400μeV at T=5K). The dependence on temperature also contrasts with the independence of transverse spin lifetimes up to ~150K.[6] Comparable lifetimes extracted from the TRA data in Fig. 5(f) suggest that carrier recombination may limit these longitudinal spin relaxation times.

Although often attributable to experimental artifacts, signals at negative time delay can correspond to an incomplete decay of spin polarization over the laser repetition interval (typically 4μs). Inspection of Figs. 8(a)-(b) indicates systematic changes in the negative delay signal with field and temperature that reveal the presence of a μs-scale decay component which accounts for the 'offset' *C* in the bi-exponential function used above. This signal represents the weighted sum of contributions from preceding pump pulses in the pulse train, each separated by the laser repetition interval. Due to practical limitations of mechanical delay lines, we are unable to directly monitor the decay of Faraday rotation on this time scale although we can obtain information on changes in the decay time by monitoring the magnitude of the negative delay signal. Figure 9(a) shows Faraday rotation data taken at fixed time delay as a function of magnetic field. The signal arising from the μs-scale decay component decreases when the field is swept toward zero, forming a roughly symmetric 'dip'. A field-induced increase in the long decay lifetime occurs over the ~20mT width of the dip which is comparable to that inferred from the data in Fig. 8(c) for the 10ns-scale component. The sign of the dip reverses with pump helicity, verifying that the negative delay signal represents a preservation of longitudinal spin polarization over microseconds. The prominence of the dip is a measure of how fully the spin polarization decays within the repetition interval. The dip decreases with increasing temperature and is nearly absent by T=40K, indicating the sensitivity of the long decay lifetime to thermal processes. As an indirect way to estimate the long lifetime component, data were taken by electronically varying the repetition interval of the laser [Fig. 9(b)]. A plot of the signal at H= -0.075T versus repetition interval roughly yields an estimate of ~20μs for the decay time. This number is significantly longer than previously measured photoluminescence lifetimes of ~ 1 μs,[15,30] perhaps suggesting that the long spin polarization is exhibited in NCs where the electron and hole have been spatially separated. The suppression of radiative recombination through such separation may increase carrier lifetimes that otherwise could set an upper limit on measured spin lifetimes. Previous studies have suggested that charge separation through Auger-induced photoionization of NCs may be an important mechanism in determining the optical emission spectrum of individual NCs[37] and energy relaxation.[38]

The exact origin for the field and temperature dependencies of the long time scale component is unclear. Here we speculate on one possible mechanism involving the hyperfine interaction with unpolarized nuclear spins.[39] An important channel for electron spin relaxation in semiconductors is based on mutual spin flips between electrons and nuclei.[40] Because many spin relaxation mechanisms in bulk semiconductors are absent in QDs, it has been suggested that spin flips with nuclei set an ultimate (although not necessarily realized) limit on the electron spin lifetime.[5] Electron spin relaxation via hyperfine interactions results in a transfer of polarization to the nuclear spin system, so that nuclear polarization well in excess of equilibrium values can be maintained with optical illumination. Once a steady state nuclear polarization has been established, further electron spin relaxation via spin exchange with nuclei is suppressed. This scenario then predicts that the changes in decay time inferred through temperature and field dependencies of the negative delay signal reflect changes in the efficiency by which the nuclear polarization can be maintained. The characteristic magnetic field width of the dips is similar in scale to dips measured in the hyperfine splitting of exciton states in GaAs QDs (~80mT) which were attributed to nuclear depolarization that occurs when the applied field is smaller than fluctuations in the electron hyperfine field.[41] Dynamic nuclear polarization (DNP) via the hyperfine interaction is a temperature dependent process whose efficiency starts to drop around 60K,[42] which is not dissimilar to the temperature dependencies observed here. Although the temperature and field scales may be comparable to previous studies, we note that efforts to directly observe DNP in CdSe NCs using the techniques in Refs. 40,42 have been unsuccessful.

Because selection rules are poorly defined in the large majority of NCs whose c-axis is oriented at a random angle, the boundaries between Faraday and Voigt geometries are blurred to the point that both electron and exciton spin precession have been observed in the Faraday geometry. Figure 10 shows data focusing on the early time dynamics where relatively weak oscillations are visible on top of the monotonically decaying signal. The frequencies of precession are identical to data taken in the Voigt geometry, and oscillation lifetimes also decrease with magnetic field. Electron precession in the Faraday geometry might be attributable to the mixture of spin-up/spin-down conduction band states that comprise the allowed exciton states[14] or result from superpositions of exciton states generated in NCs whose *c*-axis is at an angle to the field. Exciton precession seems to be sensitive to the choice of pump energy, with peak visibility occurring when $E_{pump}$ is tuned in the vicinity of the $2S_{3/2}1S_e$ state. This dependence on $E_{pump}$ is significantly different than in the Voigt geometry [Fig. 6], and might indicate that coherent superpositions of exciton spin states can be obtained upon relaxation from higher excited states. This will briefly be discussed in the Appendix.

## VI. CONCLUSIONS

We have presented systematic studies of spin dynamics under both transverse and longitudinal magnetic fields. Interpretation of the experimental data is significantly complicated by the random orientation of nanocrystals that exhibit inherent anisotropy. We have developed an assignment of the multiple frequency components visible in spin precession to electron and exciton spins. While the assignment of one frequency component to electron precession is supported by the close agreement between experimental data and theoretical calculations, interpretation of the other frequency components in terms of exciton precession remains uncertain. The quasi-spherical model discussed in the Appendix can account for several aspects of the data, but also raises new questions concerning how coherent states are optically excited and preserved under energy relaxation. Inhomogeneous dephasing in these NCs follows the simple form expected for a Gaussian distribution of *g*-factors, although the size distribution alone is insufficient to account for the observed dephasing rates. Additional contributions to the electron dephasing rate from anisotropy and surface conditions may also be important. Longitudinal spin lifetimes up to ~20μs are found to sensitively depend on applied field and temperature, exhibiting trends in both that are distinctly different than those measured for transverse spin lifetimes. The observation of spin precession in the Faraday geometry also raises questions about the nature of coherent states optically excited by circularly polarized light in these NCs.

## ACKNOWLEDGEMENTS

We are grateful to X. Peng, B. Boussert and A.P. Alivisatos for NC synthesis and collaborative efforts and to M. Poggio for his assistance. This work was supported by grants from the DARPA (MDA972-01-1-0027), the NSF (DMR-0071888) and the ONR (N00014-99-1-0077).

## APPENDIX: 'QUASI-SPHERICAL' MODEL

The lifting of degeneracy for exciton states in CdSe NCs is due in part to the wurtzite crystal structure and nonspherical shape, both of which split states in the valence band according to the projection of hole angular momentum onto the c-axis. As an example, the $A,B$ excitons in bulk CdSe result from the splitting of heavy- and light-hole states by an energy $\Delta_{cr} \sim 25 meV$. Similarly, it has been calculated that the nonspherical shape of CdSe NCs results in a splitting $D_{sh}$ $\mu$ $-m/R^2$ where $(1+m)$ is the NC aspect ratio and $R$ is the NC radius.[43] The net splitting of the $1S_{3/2}$ hole states from the intrinsic crystal structure and the NC shape is the sum of the two terms: $\Delta_{tot} = n\,\Delta_{cr} + \Delta_{sh}$, and can equal zero for certain combinations of shape and size ($n$ is a numerical factor of order 1 for CdSe NCs).[14] Figure 11(a) shows the ellipticity needed to satisfy the 'quasi-spherical' condition for the range of NC sizes studied here. Given an expected variation in $m$ of ~ ±0.2, a significant fraction of NCs within the ensemble may satisfy this condition. The exciton $g$-factors associated with the $1S_{3/2}1S_e$ states in quasi-spherical NCs may be largely isotropic due to the lack of any preferred axis with respect to the applied magnetic field. The energies of exciton fine structure states are solely determined by the electron-hole exchange interaction which splits the initially eightfold degenerate exciton into two levels which are five- and three-fold degenerate according to $\boldsymbol{J}^{tot} = \boldsymbol{J}^{h} \pm \boldsymbol{S}$ [14] [Fig. 11(b)].

The relation for $g_{exc}$ in Eq. (3) can be calculated using expressions for the exciton wavefunctions from Ref. 14 evaluated in quasi-spherical NCs:

$$\Psi_1^L(r_e, r_h) = +i\frac{\sqrt{3}}{2} y_{\uparrow,1/2}(r_e, r_h) + \frac{1}{2} y_{\downarrow,3/2}(r_e, r_h)$$

$$\Psi_{-1}^L(r_e, r_h) = +\frac{i}{2} y_{\uparrow,-3/2}(r_e, r_h) + \frac{\sqrt{3}}{2} y_{\downarrow,-1/2}(r_e, r_h).$$

The $y$ 's here represent $1S_{3/2}1S_e$ excitons with electron and hole angular momenta given by the subscripts. Expressed in the basis of state vectors $y$, the Hamiltonian matrix which describes the Zeeman splitting in an applied magnetic field is (Eq. 30 in Ref. 14):

$$\hat{H}_{Zeeman} = m_B H_z \begin{bmatrix} a & & & & & & & \\ & b & & & & & & \\ & & g & & 0 & & & \\ & & & d & & & & \\ & & & & -d & & & \\ & & 0 & & & -g & & \\ & & & & & & -b & \\ & & & & & & & -a \end{bmatrix} \text{ where } \begin{aligned} a &= (g_e - 3g_h)/2, \\ b &= (g_e - g_h)/2, \\ g &= (g_e + g_h)/2, \\ d &= (g_e + 3g_h)/2. \end{aligned}$$

The effective exciton $g$-factor in quasi-spherical NCs is then calculated from the relation:

$$g_{exc}^{1^L} m_B H_z = \left\langle \Psi_{+1}^L \middle| \hat{H}_{Zeeman} \middle| \Psi_{+1}^L \right\rangle - \left\langle \Psi_{-1}^L \middle| \hat{H}_{Zeeman} \middle| \Psi_{-1}^L \right\rangle .$$

Effective $g$-factors for the different fine structure states calculated using this procedure are:

$$\begin{aligned} g_{exc}^{1^L} &= (g_e - 3g_h)/2 \\ g_{exc}^{1^U} &= -(g_e + 5g_h)/2 \\ g_{exc}^{2} &= (g_e - 3g_h). \end{aligned}$$

The hole $g$-factor $g_h$ (= -0.73) was treated as a fitting parameter for the calculation of $g_{exc}$ in Fig. 3 as there are no reliable experimental data for this quantity to our knowledge. While the calculated value of $g_{exc}^{1^L}$ agrees with the experimental data for the higher $g$-factor components in the entire sample set (using the same value of $g_h$), the other excitonic $g$-factors in the 57Å NCs can't be accounted for with $g_{exc}^{1^U}$, or $g_{exc}^{2}$.

Because the $\pm 1^L$ states in quasi-spherical NCs are optically inactive,[14] it is expected that the initial excitation instead involves either the $\pm 1^U$ level or higher excited states (i.e. $2S_{3/2}1S_e$ etc.). These higher-lying states become increasingly inaccessible with decreasing $E_{pump}$, which may account for the decrease in exciton precession observed in Fig. 6. As mentioned previously, the timescale for relaxation among the fine structure states is not well known. The scenario presented here relies on the coherent relaxation of higher excited states to the $\pm 1^L$ level, which then exhibits a spin splitting corresponding to $g_{exc}^{1^L}$. This is currently an open question that requires further study. Exciton precession in the Faraday geometry may be explained with similar arguments. A priori, exciton precession in this geometry is unexpected, since well-defined optical selection rules for the $1S_{3/2}1S_e$ state in quasi-spherical NCs dictate that circularly polarized light directly excites a particular spin eigenstate rather than superpositions of eigenstates. Such restrictions don't hold for excitation into higher excited states; the exciton might then relax coherently into the ground state and precess as usual.

Definitive understanding of exciton precession in these NCs is difficult given the uncertain nature of the energy spectrum and optical selection rules over the ensemble of NCs. While the quasi-spherical model addresses this difficulty, it raises a host of additional questions related to the nature of coherent excitation. Analogous studies of spin dynamics in related materials, such as CdTe NCs (which have a cubic crystal structure) may yield further insight into these problems.

TABLE 1

Compilation of $g$-factors observed in CdSe NCs from spin precession measurements taken at T=5K.

| | 22Å | 25Å | 25Å Core/shell | 40Å | 40Å Core/shell | 57Å | 80Å |
|---|---|---|---|---|---|---|---|
| $g_e$ | 1.63±0.01 | 1.565±0.002 | 1.631±0.004 | 1.30±0.01 | 1.458±0.004 | 1.138±0.006 | 1.014±0.006 |
| $g_2$ | - | 1.83±0.01 | 1.85±0.01 | 1.72±0.01 | 1.78±0.01 | 1.636±0.007 | 1.64±0.02 |
| $g_3$ | - | - | - | - | - | 1.734±.004 | - |
| $g_4$ | - | - | - | - | - | 1.46±0.05 | - |

# FIGURE CAPTIONS

Figure 1. (a) Absorption spectrum of 57Å NCs at T=6K. Ground and excited energies can be extracted from the respective components in a multi-Gaussian fit (dotted lines). (b) Unresolved fine structure states within the NC exciton ground state (degeneracies are indicated in parentheses).

Figure 2. (a)-(e) Time-resolved Faraday rotation in the Voigt geometry. T=5K, H=4T. Laser parameters are: 80Å: $E_{pump}$=$E_{probe}$= 2.00eV; 57Å: $E_{pump}$=2.07eV, $E_{probe}$=2.10eV; 40Å: $E_{pump}$= $E_{probe}$= 2.14eV; 25Å: $E_{pump}$=$E_{probe}$=2.48eV. (e) $E_{pump}$= 2.175eV, $E_{probe}$ = 2.019eV or 2.039eV, $P_{pump}$= 0.5mW, $P_{probe}$=20μW. (f) Fourier transform of data in (d) showing two distinct precession frequencies.

Figure 3. Compilation of $g$-factors measured in CdSe NCs. The open symbols are from analysis of spin precession frequencies, while the two closed symbols are extracted from spin splittings in magnetoabsorption spectra. The solid and dotted lines are calculations of the size-dependent electron $g$-factor using Eq. (2) and Ref. 7(b) respectively. The dashed line is a calculation of an exciton $g$-factor using Eq. (3) with $g_h$=-0.73.

Figure 4. Calculations of exciton fine structure states versus size, adapted from [14]. (a) Spherically shaped NCs ($m$ =0). (b) Prolate NCs ($m$ =0.28).

Figure 5. (a) Absorption spectrum of 57Å NCs with pump energies in (b)-(d) marked by arrows for reference. (b)-(d) Pump-induced changes in absorption taken at fixed time delay with either a co- or counter-circularly polarized probe. Use of the white light continuum here allows broad spectral coverage of $DaL$. $P_{pump}$=0.6mW. (b) $E_{pump}$=1.968eV. (c) $E_{pump}$=2.138eV. (d) $E_{pump}$=2.296eV. (e) Time scan showing ultrafast energy relaxation. $E_{pump}$=2.296eV, $E_{probe}$=2.109eV, $P_{pump}$=0.5mW, $P_{probe}$=30μW. (f) Comparison of decays in $DaL$ at T=6 and 70K. $E_{pump}$=2.142eV, $E_{probe}$=2.105eV $P_{pump}$=0.9mW, $P_{probe}$=100 μW.

Figure 6. Dependence of TRFR on $E_{pump}$ in 57Å NCs. $P_{pump}$=0.5mW, $P_{probe}$=30μW, $E_{probe}$=2.109eV, H=4T, T=6K. The gray bar highlights the changing beat pattern between exciton and electron precession. Right: Corresponding FFTs show an increase in exciton precession with $E_{pump}$.

Figure 7. Plots of FFT linewidth ($D_f$) versus field reveal inhomogeneous dephasing in CdSe NCs. (a) Electron and exciton dephasing in 80Å NCs. Inset: Peaks in the FFT and Gaussian fits (solid line) of central regions used to extract $D_f$. (b) Comparison of dephasing in 40Å core and core/shell NCs. (c) Size-dependence of the electron dephasing rate. Inset: Variance in $g_e$ calculated from spin precession data (closed symbols) and expected from size distributions of 5% (open squares), 10% (left triangles), and 15% (right triangle). The lines are guides to the eye.

Figure 8. TRFR in 57Å NCs taken in the Faraday geometry with $P_{pump}$=1.0mW, $P_{probe}$=120μW, $E_{pump}$=2.138eV, $E_{probe}$=2.102eV. (a) The decay time of longitudinal spin polarization increases with field from H=0T (bottom) to H=1T (top). (b) Temperature dependence taken at fixed field. (c)-(d) Longitudinal spin lifetimes extracted from fits to (a)-(b) assuming a bi-exponential decay. Systematic changes with temperature and field in the signal at negative delay reveal an additional μs-scale decay component.

Figure 9. (a) Changes in the μs-scale decay time of longitudinal spin polarization are inferred from the field and temperature dependence of the negative delay TRFR signal. $E_{pump}$=2.138eV, $E_{probe}$=2.102eV, $P_{pump}$=0.9mW, $P_{probe}$=100μW. (b) Variation of the laser repetition interval provides a rough estimate of ~20μs for the decay time. T=5K. The pump power was varied to keep the energy per pulse constant: $P_{pump}$=(2.1mW)(3.797μs/rep interval), $P_{probe}$=170μW.

Figure 10. Spin precession in the Faraday geometry with varying pump energy. FFTs were calculated after subtracting the monotonically-decaying background to reduce the baseline signal and improve peak visibility. Data were taken with $E_{probe}$=2.175eV, $P_{pump}$=2mW, $P_{probe}$=40 μW.

Figure 11. (a) Calculation of ellipticity versus size needed to satisfy the quasi-spherical condition. (b) Calculation of five- and three-fold degenerate exciton fine structure levels for quasi-spherical NCs.

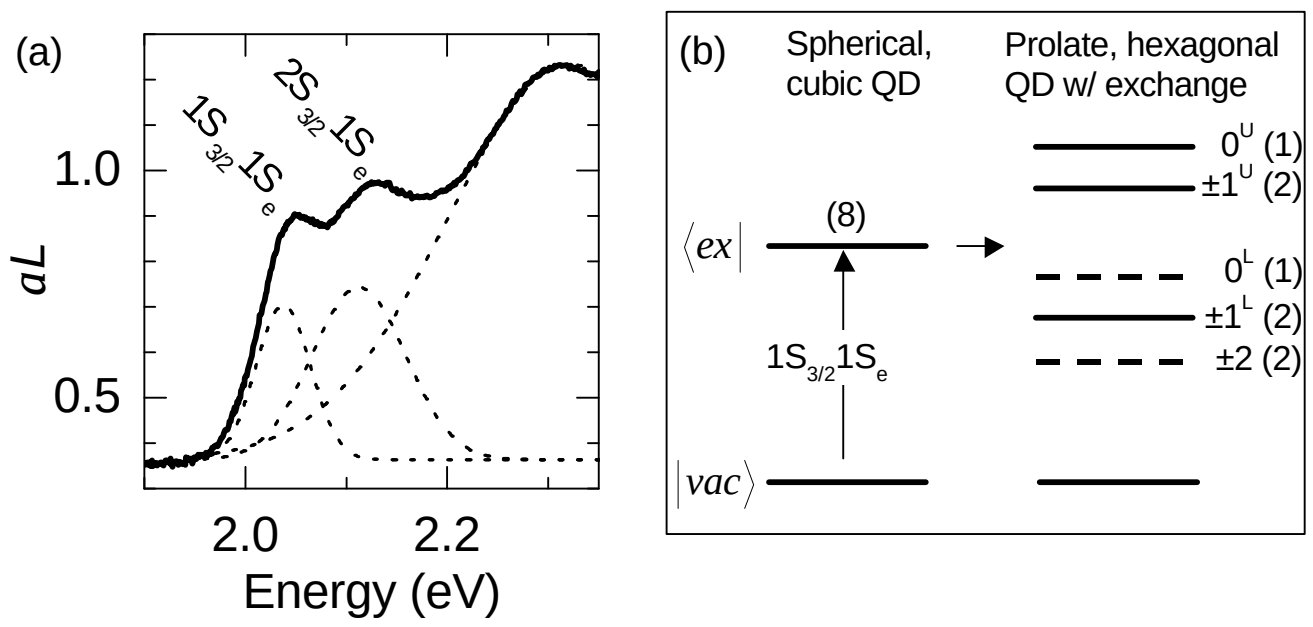

Gupta et al., Figure 1

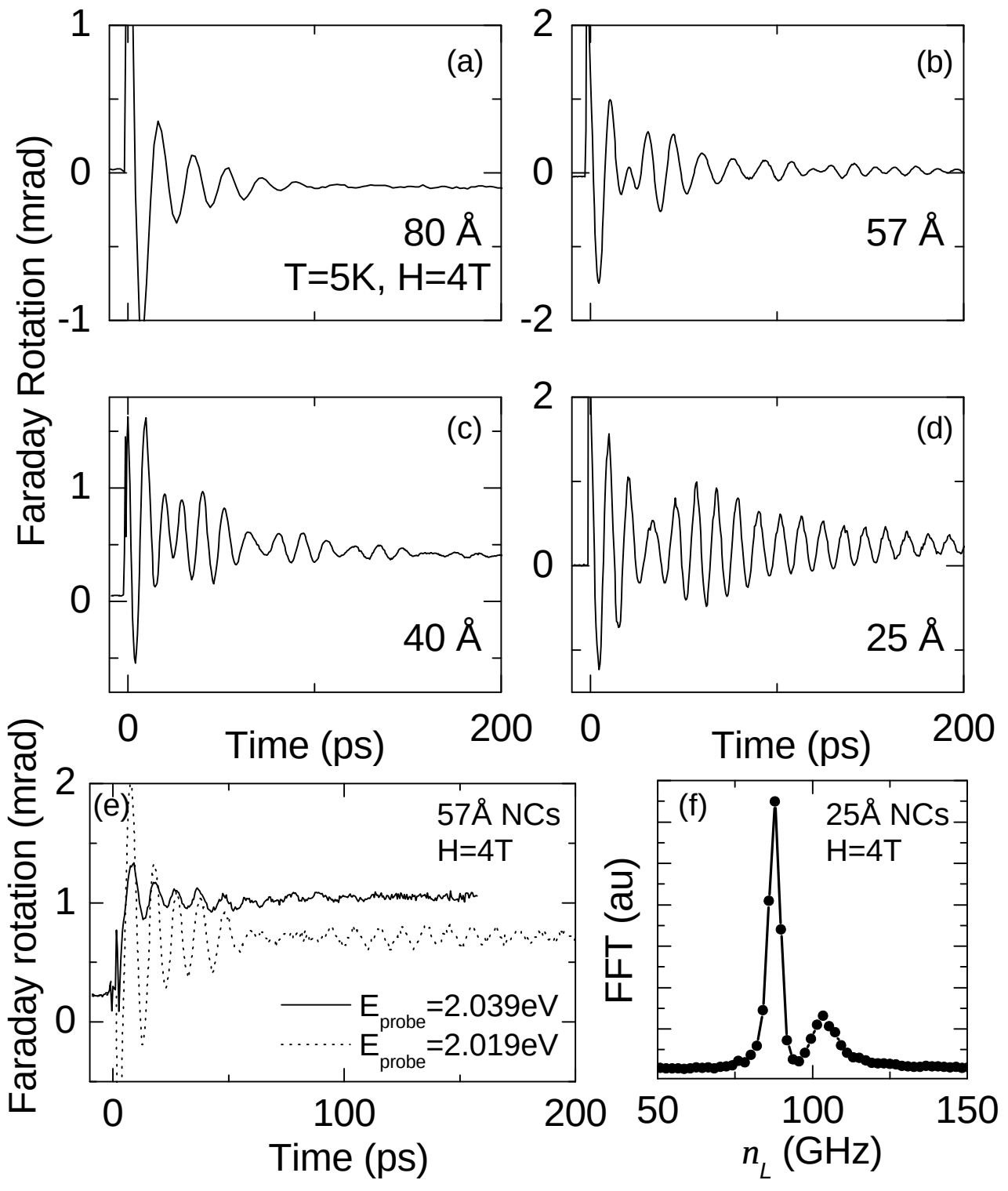

Gupta et al., Figure 2

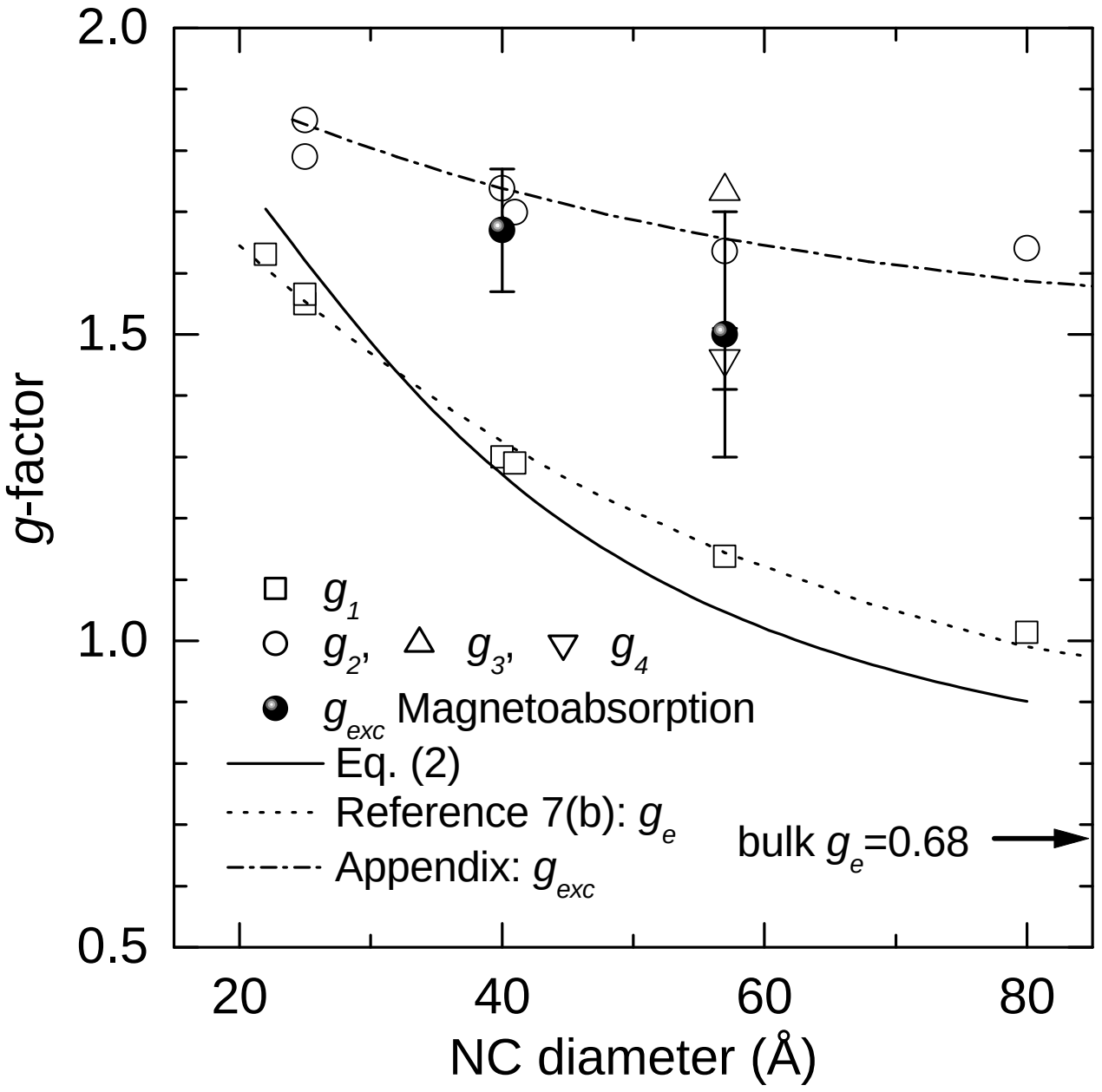

Gupta et al., Figure 3

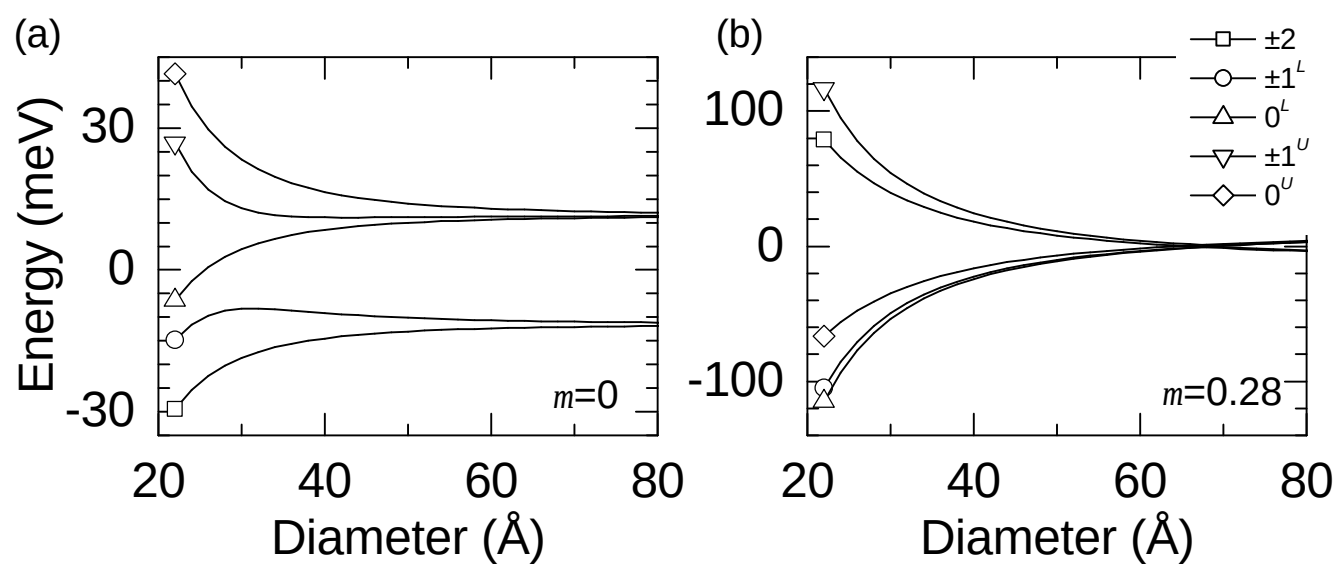

Gupta et al., Figure 4

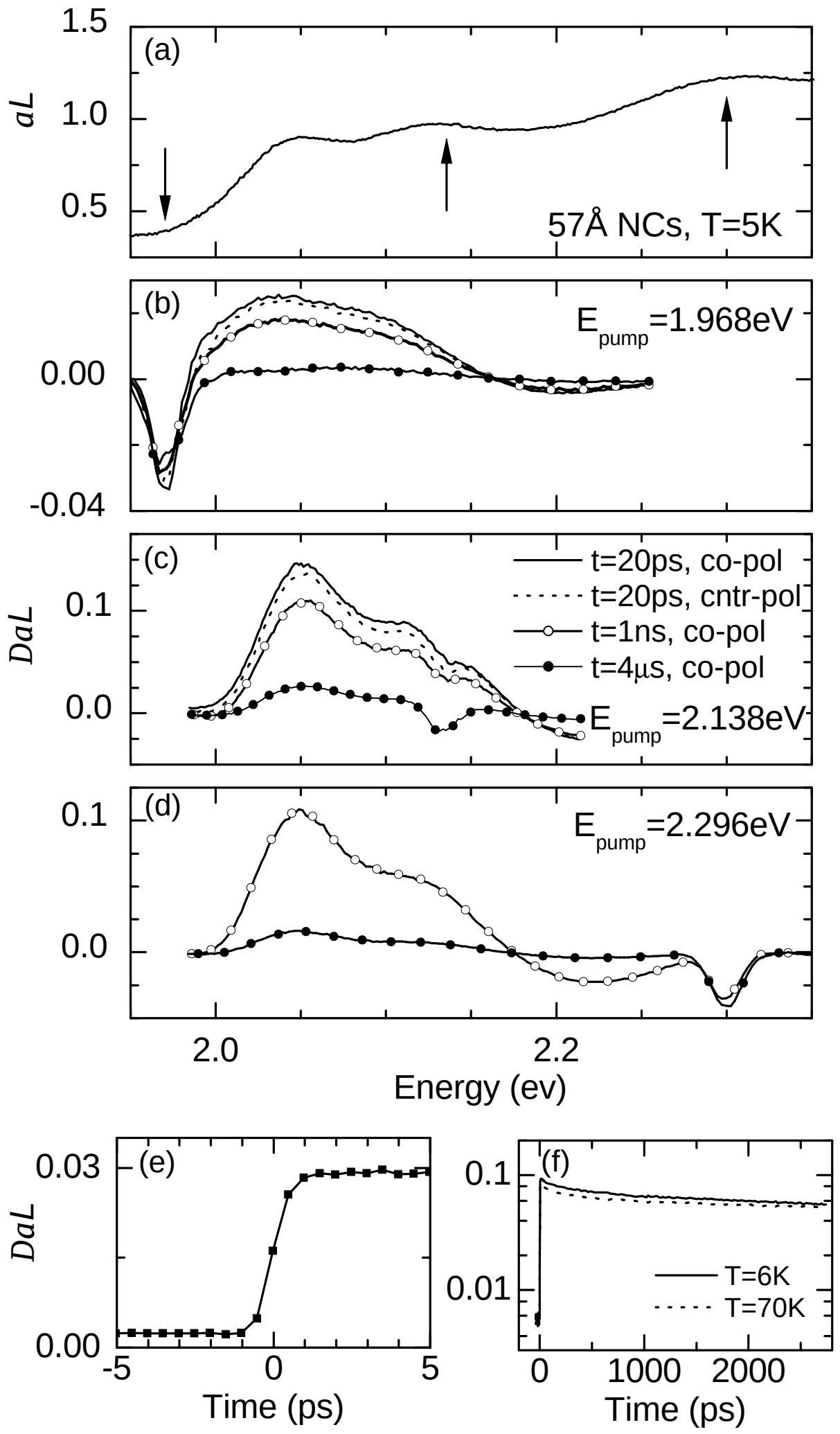

Gupta et al., Figure 5

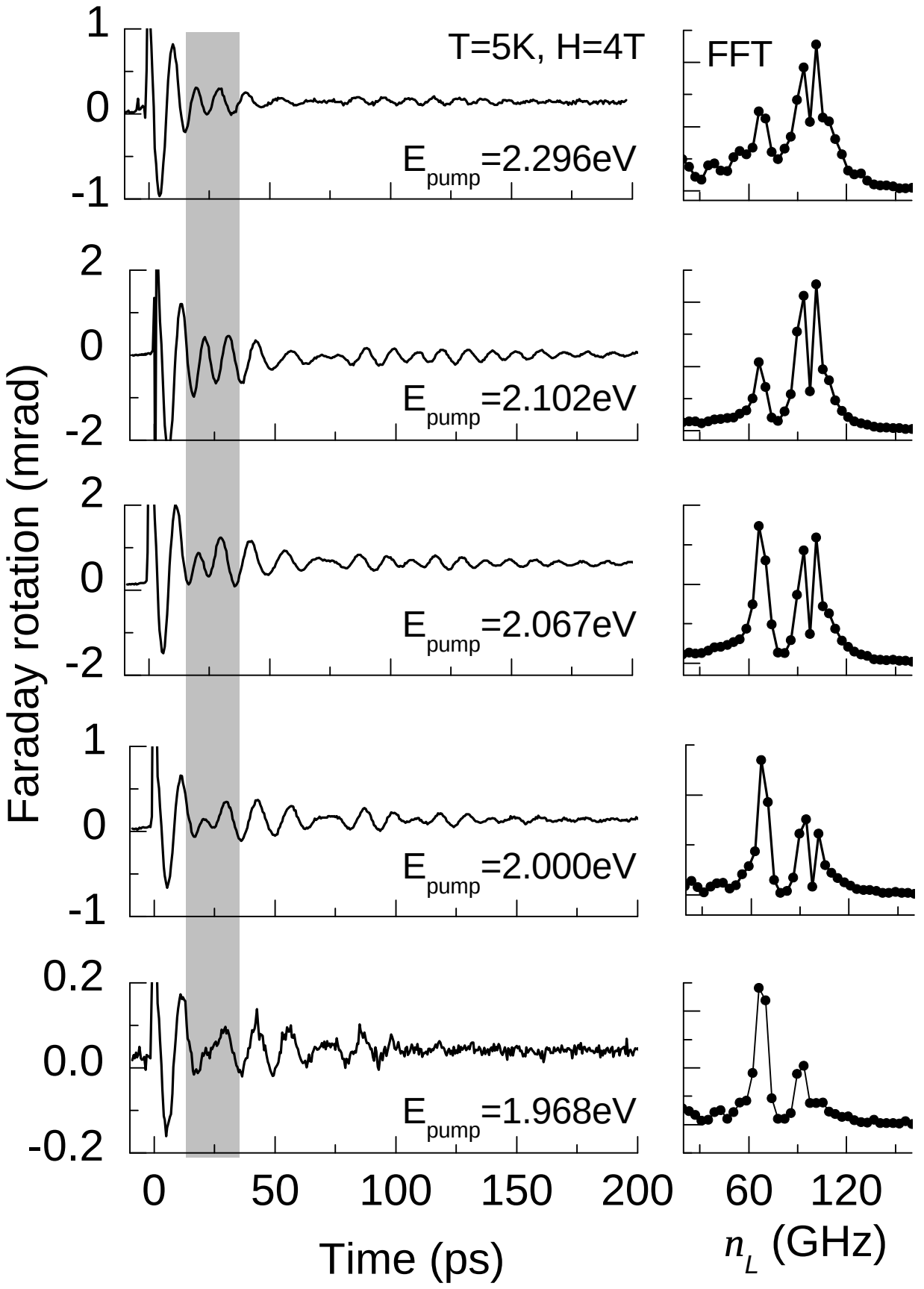

Gupta et al., Figure 6

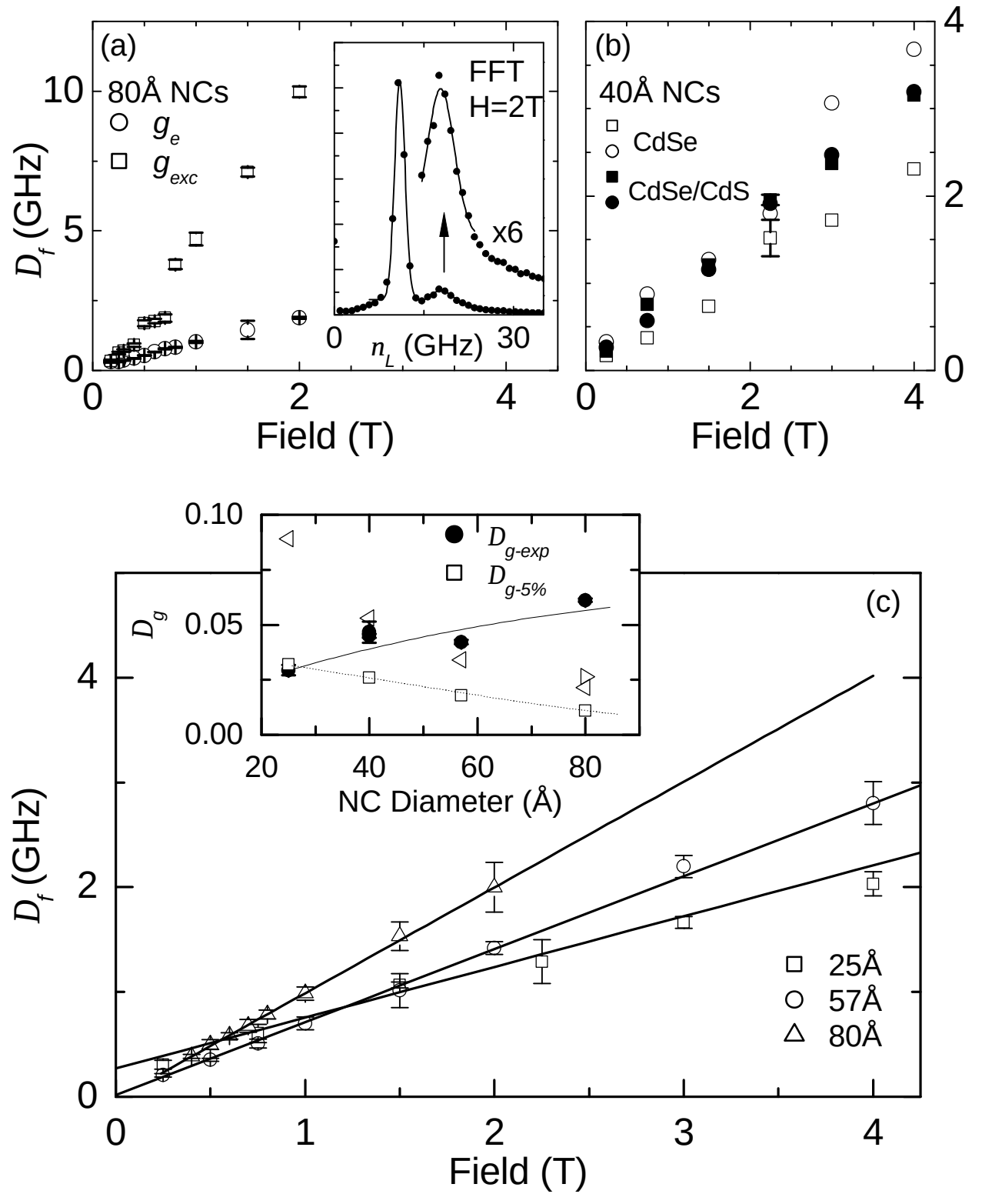

Gupta et al., Figure 7

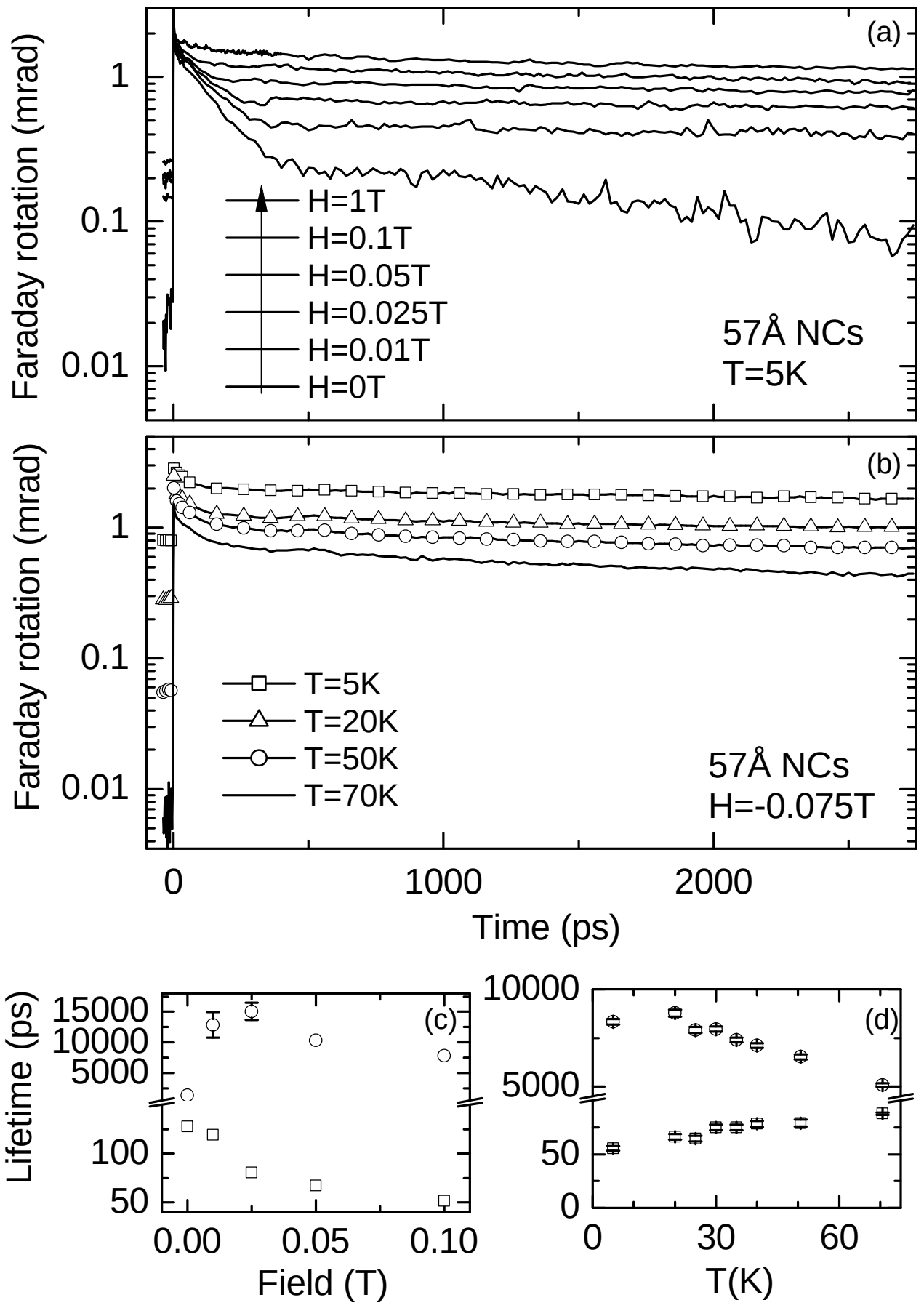

Gupta et al., Figure 8

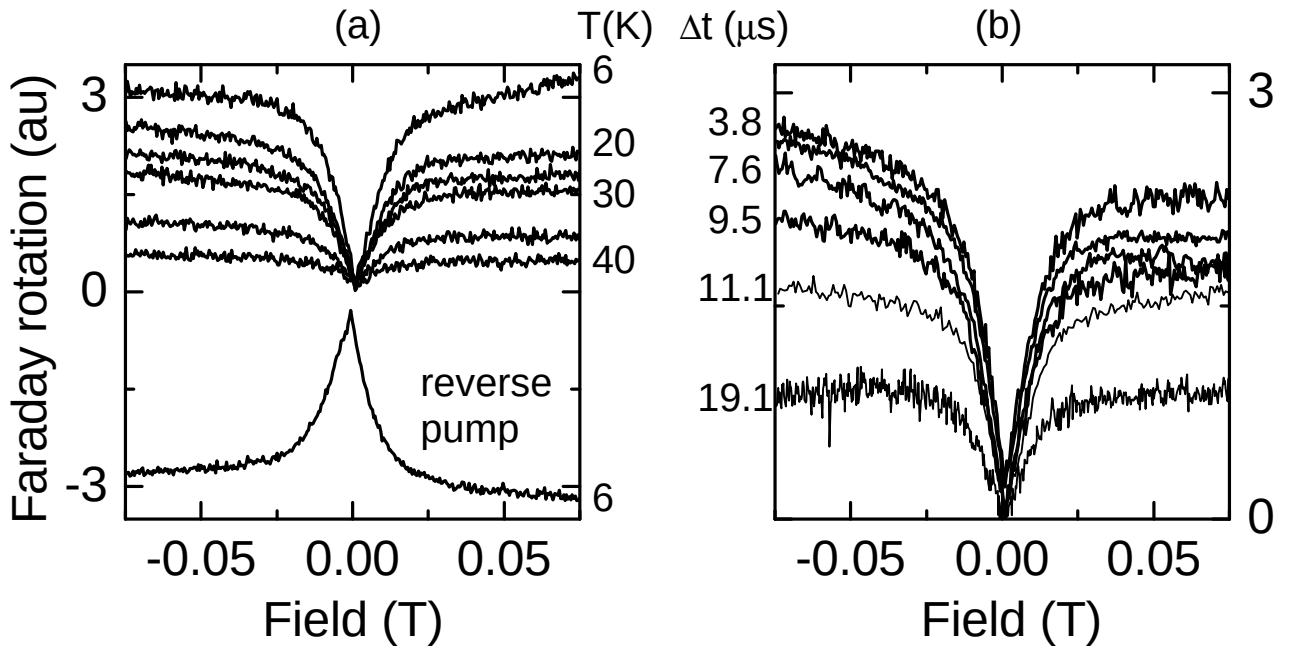

Gupta et al., Figure 9

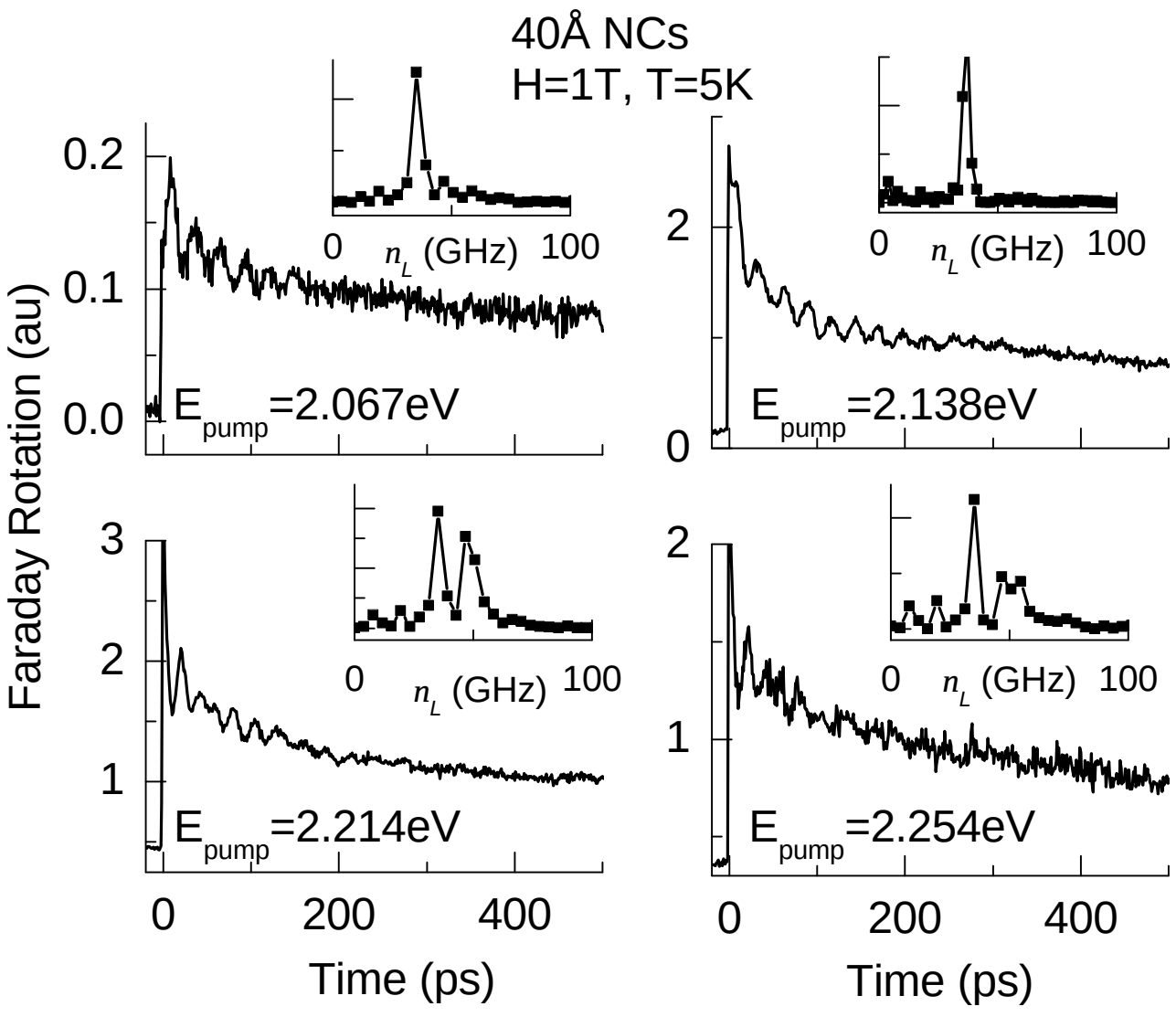

Gupta et al., Figure 10

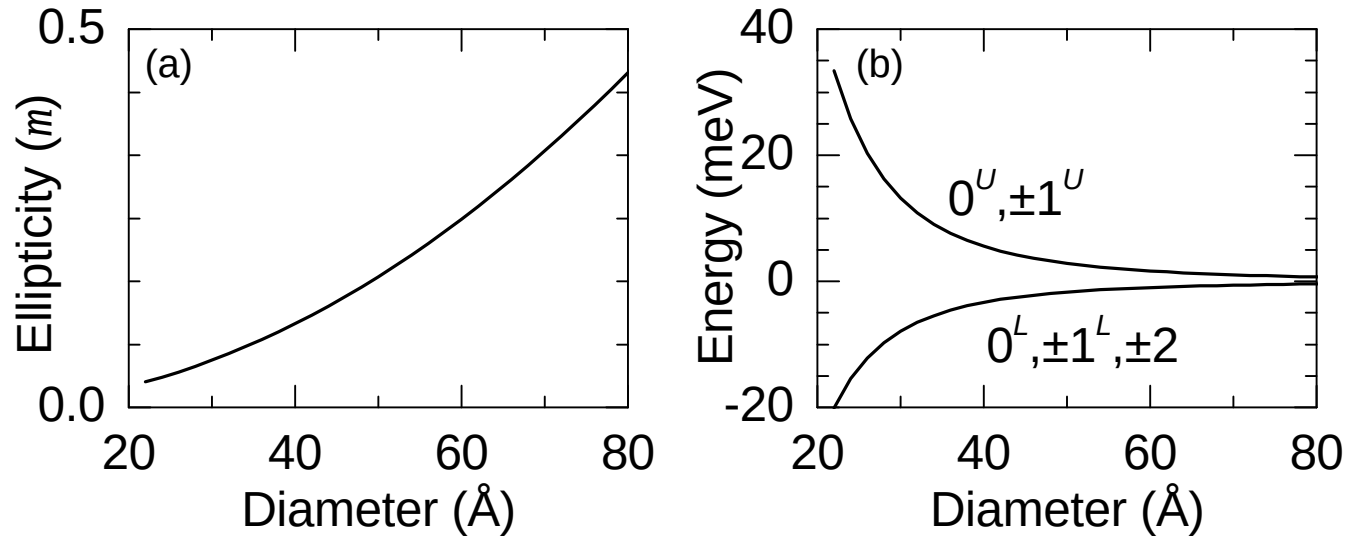

Gupta et al., Figure 11